\documentclass[a4paper, amsfonts, amssymb, amsmath, reprint, showkeys, nofootinbib, twoside]{revtex4-1}
\usepackage[english]{babel}
\usepackage[utf8]{inputenc}
\usepackage{epsfig}
\usepackage{comment}
\usepackage{booktabs}
\usepackage{float}
\usepackage{subcaption}
\usepackage[colorinlistoftodos, color=green!40, prependcaption]{todonotes}
\usepackage[pdftex, pdftitle={Article}, pdfauthor={Author}]{hyperref} 
\begin{document}
\title{The isostructural $\alpha-\gamma$ phase transition in cerium from the perspective of meta-generalized gradient approximations }

\author{Ashesh Giri, Chandra Shahi and Adrienn Ruzsinszky}
    \email[Correspondence email address: ]{aruzsin@tulane.edu}
    \affiliation{Department of Physics and Engineering Physics, Tulane University, New Orleans, LA 70118 }

\date{\today} 

\begin{abstract}
\noindent
Meta-generalized gradient approximations (meta-GGAs) on the third rung of the functional hierarchy are gaining increasing relevance for the electronic structure. Meta-GGAs are constructed from numerous ingredients including the orbital kinetic energy density that make them more flexible than generalized gradient approximations (GGAs) including the heavily used PBE-GGA. Still, most meta-GGAs cope with the expected limitations of a semilocal density functional when band gaps or localization of electrons are needed. On the other hand, meta-GGAs are implicit functionals of the orbitals. This feature resembles hybrid density functionals with exact exchange. Efforts in recent years demonstrate that some meta-GGAs can rise beyond the accuracy of semilocal approximation when band gaps are computed. Cerium is an ideal testbed to challenge some recent meta-GGAs. Cerium shows an isostructural $\alpha-\gamma$ phase transition with delocalized and localized f electrons in each phase, respectively. Since the phonon entropy term was found negligible in the $\alpha-\gamma$ phase transition of cerium by accurate experiments, all changes in the transition are driven by electronic correlation. The correlation of f electron systems is hardly captured by semilocal approximations but the recent LAK meta-GGA with ultranonlocality steps out of the framework of conventional semilocal density functionals and delivers spectacular accuracy for the phase transition of cerium. LAK and further meta-GGAs inspired by the success of LAK can open a forefront of meta-GGAs for quantum materials with localized electrons. 
\end{abstract}


\maketitle

A first principles characterization of f-electron materials has been known as a high barrier for semilocal density functional approximations without exact exchange or Hubbard correction. The partially occupied f states usually call for a self-interaction correction, while the filled spd states can be described accurately at the semilocal level of density functionals. Cerium encompasses all these challenges in the electronic structure. Cerium can undergo an isostructural phase transition between the two paramagnetic $\alpha$ and $\gamma$ phases following a volume collapse when the cerium turns to the $\alpha$ structure. The $\gamma$ phase displays localized magnetic moments with the magnetic susceptibility satisfying the Curie-Weiss law \cite{koskenmaki1978cerium}.\\

The phase transition of cerium is intrinsically coupled with delocalization and localization of the f electrons. In the $\gamma$ phase, electrons are localized, consequently leading to localized magnetic moments. In contrast, f electrons in the $\alpha$ phase (smaller volume) are delocalized, resulting in typical Pauli paramagnetic behavior. A comprehensive theory should aim for a simultaneous characterization of both $\alpha$ and $\gamma$ phases with delocalized and localized f electrons. Numerous previous studies confirm that this goal is not realistic at the level of semilocal approximations \cite{eryigit2022gamma, svane1994electronic, casadei2012density, szotek1994self}. With different explanations of the driving mechanism, the role of a comprehensive and unbiased theory model becomes highly relevant. Furthermore, we want our model to be parameter free.\\ 

Semilocal and hybrid density functional approximations largely differ when describing the localized f electrons. Exact exchange, self-interaction correction or Hubbard-corrected DFT capture localization at the expense of increased computational cost or additional empiricism. The local density approximation (LDA) \cite{kohn1965self} and the Perdew-Burke- Ernzerhof (PBE) generalized gradient approximation \cite{perdew1996generalized} capture only the alpha phase, while the recent strongly constrained and appropriately normed SCAN \cite{sun2015strongly} meta-generalized gradient approximation (meta-GGA) shows a minimum energy for $\alpha$ and a shoulder for the $\gamma$ phase. The magnetic moment for $\gamma$-Ce using SCAN is in good agreement with that of the global hybrid PBE0 hybrid functional. \\

Meta-GGAs are semilocal functionals of the occupied orbitals but nonlocal functionals of the density. In this sense, meta-GGA approximations are regarded as methods based on the density matrix instead of the density itself. Consequently, meta-GGA density functionals, in principle, show some analogy with hybrid density functionals.
Still, meta-GGAs often produce a biased accuracy for structural or electronic properties. Meta-GGA functionals add the extra kinetic energy density or the reduced Laplacian as ingredients beyond the GGA level. Each choice is a great additional degree of freedom to cover a broad range of physics, including improved fundamental band gaps, charge transfer or magnetic moments compared to GGA approximations.\\ 

The ambitious dream is to keep all the exact constraints and appropriate norms of meta-GGAs as SCAN or $\mathrm{r^2SCAN}$ \cite{furness2020accurate} were constructed to do, while adding new norms so that electron localization is accounted for. While this remains to be seen, recent meta-GGAs have had undoubtedly great successes in reducing self-interaction error (SIE) and capturing electron localization. An improvement for band gaps has already been noticed for SCAN and $\mathrm{r^2SCAN}$. While the more recent TASK \cite{aschebrock2019ultranonlocality} and mTASK \cite{neupane2021opening}are great choices for band gaps, their structural properties significantly worsen when compared to SCAN or $\mathrm{r^2SCAN}$. The very recent LAK meta-GGA \cite{lebeda2024balancing} applies a different construction scheme that elevates it close to becoming an ``ultranonlocal" meta-GGA. \\ 

The isostructural phase transition of cerium is an ideal testbed to demonstrate the unexplored flexibility of meta-GGA approximations. \\

Semilocal DFAs can be written in the generic form as:\\
\begin{equation}
    \centering
    E_{xc}[n]=\int d^3r n\epsilon_{xc}(n,\nabla n,\nabla^2n,\tau)
\end{equation}

In LDA \cite{kohn1965self}, the only ingredient is the electron density, n(r). Generalized gradient approximations (GGA) \cite{perdew1996generalized} add the electron density gradient, $\nabla n$. Meta-GGAs \cite{sun2015strongly,furness2020accurate,kaplan2022laplacian, lebeda2024balancing,mejia2018deorbitalized,aschebrock2019ultranonlocality,neupane2021opening}further include the Laplacian of the electron density, $\nabla^2 n$, and/or the kinetic energy density, $\tau=\frac{1}{2} \sum_{i=1}^{N_{occ}}|\nabla\phi_i|^2
$, this latter functional constructed from the Kohn-Sham orbitals $\phi_i$. The more ingredients semilocal functionals have, the more exact constraints they can satisfy.\\

The most dominant group of meta-GGAs are built upon the kinetic energy density ingredient. The most representative examples are SCAN, $\mathrm{r^2SCAN}$, the more recent TASK or mTASK and LAK. The kinetic energy density makes these meta-GGAs more nonlocal than others with the Laplacian ingredient only. SCAN and $\mathrm{r^2SCAN}$ are designed to yield accurate equilibrium properties. $\mathrm{r^2SCAN}$ is a numerically more practical realization of SCAN. It satisfies all 17 exact constraints and norms, as SCAN does, thereby preserving accurate equilibrium lattice constants, cohesive energies and lattice dynamics. 

Some of the meta-GGAs such as TASK, mTASK, and LAK can demonstrate the so-called ultranonlocal behavior as a feature of the exchange-correlation potential. Derivative discontinuity is not found for periodic solids but an ultranonlocal meta-GGA within the generalized Kohn-Sham approach can achieve increased band gaps that resemble approximations with exact exchange.  Ultranonlocality also becomes crucial at charge transfer.\\

The ultranonlocality is better explained within the more intuitive form of meta-GGAs that tells how strongly they differ from LDA:\\
\begin{equation}
    \centering
    E_{x}[n]=A_x\int d^3rn^\frac{4}{3}F_x(s,\alpha)
\end{equation}

where $A_{x} = -(\frac{3e^2}{4})(\frac{3}{\pi})^\frac{1}{3}$, and $F_x(s,\alpha)$ is the exchange enhancement factor with the ingredients of $s=|\nabla n|/(2(3\pi^2)^\frac{1}{3}n^\frac{4}{3})$ the reduced gradient and $\alpha =\frac{\tau-\tau^W}{\tau^{unif}}$. $\tau^W=\frac{|\nabla n|^2}{8n}$, $\tau^{unif} =(\frac{3}{10})(3\pi^2)^\frac{2}{3}n^\frac{5}{3}$. $\alpha$ recognizes two regions, the iso-orbital limit ($\alpha = 0$) and the uniform electron gas ($\alpha = 1$) used in numerous meta-GGAs. The $\alpha$ ingredient appears in SCAN and $\mathrm{r^2SCAN}$ as well in the more recent TASK and mTASK approximations and LAK.\\

The slope of the enhancement factor as a function of $\alpha$ reveals the information on the degree of ultranonlocality of meta-GGAs. The condition of ultranonlocality is

\begin{equation}
\centering
    \frac{\partial F(s,\alpha)}{\partial\alpha}<0
\end{equation}
or equivalently
\begin{equation}
    \centering
    \frac{\partial\epsilon_x}{\partial\alpha}>0
\end{equation}
where this latter constraint is part of the functional derivative that distinguishes meta-GGA exchange-correlation potentials from GGAs.\\

The very recent LAK meta-GGA also demonstrates the ultranonlocality feature. But the construction scheme of LAK is different from its predecessors. LAK makes an adjustment of the relative contributions from the density gradient and the kinetic energy density in the gradient expansion of the exchange-correlation energy. This adjustment creates a delicate balance between ground state energetic and ultranonlocal behavior.\\

Cerium is a metal. Metals show perfect long-range screening. Consequently, the exchange-correlation energy of metals is only weakly nonlocal, at least in solids with \textit{3d} electrons. While $\tau$ is a great choice as an ingredient for insulators, it adds too much of the nonlocality to metals. In many examples orbital-free meta-GGAs become better choices for metals than meta-GGAs with orbital ingredients. Orbital-free meta-GGAs replace the analytic expression for $\tau$ with an approximate form that itself is constrained to recover exact constraints. Most recent de-orbitalized meta-GGAs such as SCAN-L \cite{mejia2018deorbitalized} or $\mathrm{r^2SCAN}$-L \cite{mejia2020meta} and OFR2\cite{kaplan2022laplacian} return improved magnetic moments in 3d metals compared to the overestimated values by SCAN or $\mathrm{r^2SCAN}$.\\

At zero temperature, a double minimum in the total energy versus volume curve is a direct indication of the phase transition. First-principles calculations without exact exchange or Hubbard correction have so far been unable to produce a double minimum.  Hybrid functionals \cite{casadei2012density} incorrectly put the energy of the $\gamma$ phase below that of the $\alpha$ phase. Local or semilocal (LDA or GGA) functionals make the f electrons always delocalized, due to the strong self-interaction error (SIE) of the approximate functionals.\\  

The success \cite{casadei2012density} of capturing the double minimum with hybrid density functionals at zero temperature provides some support for the Mott mechanism \cite{mott} with localized vs. delocalized electrons in the $\gamma$ and $\alpha$ phases, respectively. Exact exchange can properly capture electron localization. The correlation energy density of the meta-GGAs described above is constructed to be SIE-free. The exchange energy density is not SIE free but achieves a significant mitigation of SIE, as supported by numerical results. The Mott picture \cite{johansson1974alpha} ignores the contribution of spd electrons, however. The Kondo picture \cite{allen1982kondo} localizes the f electrons in both phases, and f electrons can hybridize with spd, but not with other f electrons. The phase transition is attributed to a volume change that can modify the screening of the localized moments of the spd electrons. The ingredients s and $\tau$ make meta-GGAs flexible enough to adapt to the change in screening, therefore the meta-GGA description of the phase transition in Ce can be easily reconciliated with the Kondo picture as well. But we can suspect that both $\gamma$ and $\alpha$ phase are strongly correlated \cite{strongcorr}.\\

The equilibrium structural parameters of bulk cerium phases were obtained by fitting the energy–volume (E–V) data to the Birch-Murnaghan equation of state\cite{murnaghan1944compressibility}. The E–V data were generated using the Vienna Ab initio Simulation Package (\textsc{vasp})\cite{kresse1996efficient}. Unless otherwise specified, spin-polarized calculations, with an initial magnetic moment of 3.0 $\mu$B, were performed using the LAK, OFR2 and $\mathrm{r^2SCAN}$ meta-GGA functionals. A plane-wave kinetic energy cutoff of 520 eV was employed, and the Brillouin zone was sampled using a $\Gamma$-centered k-point mesh of 24x24x24. The standard \textsc{vasp} PAW potential (Ce\_3) was used for cerium, in which the 4f electrons are explicitly treated as valence states.\\

The LAK meta-GGA spectacularly stands out against other {methods: it predicts not only the double minimum and the magnetic transition but also the correct phase ordering. 
Figure 1 shows the total energy of bulk Ce under isotropic pressure as the lattice constant decreases from 5.3 \AA\ to 4.6 \AA\ for the LAK functional. The spin-unpolarized calculation only gives the minimum energy for the high-pressure ($\alpha$) phase. However, when we performed spin-polarized calculations, we obtained the double minimum in the  the total energy curve, indicating the existence of both phases ($\alpha$ and $\gamma$). The values obtained for the lattice parameter for the $\alpha$ and $\gamma$ phases are 4.87 \AA\ and 5.14 \AA, which agree with the experimentally obtained results.
\begin{figure}[h]
    \centering
    \includegraphics[width=0.92\linewidth]{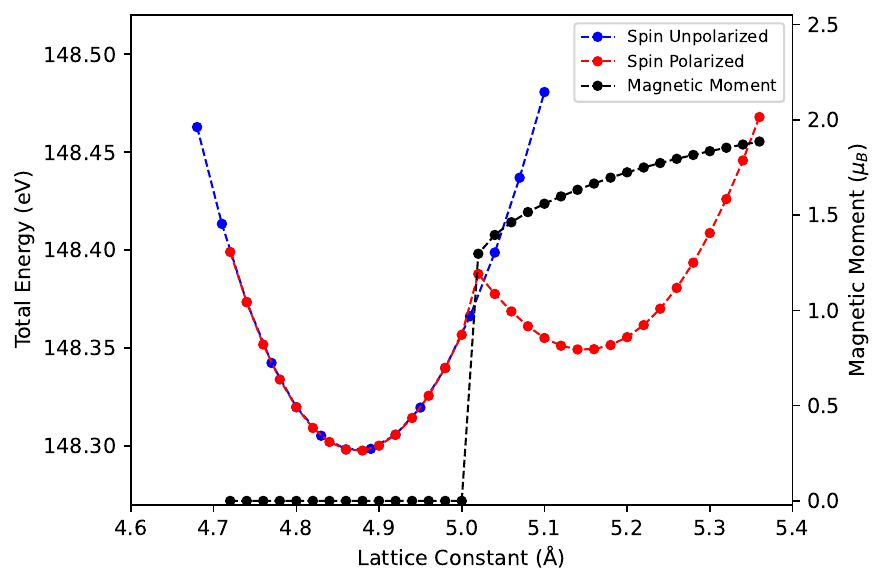}
    \caption{The LAK total energy and magnetic moment of bulk Ce as a function of the lattice constant. The blue curve from the spin-unpolarized calculation perfectly matches that obtained from a spin-polarized calculation for the $\alpha$ phase}
    \label{fig1}
\end{figure}
We also performed similar calculations with OFR2, and $\mathrm{r^2SCAN}$. Like other semilocal functionals, reported in previous studies, they only predict the $\alpha$ phase. 
The obtained values of lattice constants are 4.62 \AA\ and 4.65 \AA\ for OFR2 and {$\mathrm{r^2SCAN}$, respectively. In Table \ref{tab:1}, we summarize our results from this work and compare them with other literature values. LDA and GGA only give the $\alpha$ phase, and both functionals highly underestimate the value of the lattice parameter. The non-self-consistent calculation using exact exchange plus random phase approximation evaluated at the PBE0 orbitals was found \cite{casadei2012density} to give both phases, but the values obtained from the lattice parameters do not match the experimental values at room temperature. However, the lattice constants computed from LAK at zero temperature are in close agreement with the experimental values at room temperature. Unlike other semilocal functionals, the LAK functional is able to explain this phase transition because it has more nonlocality than other semilocal DFA's.\\
\begin{table}
    \centering
    \renewcommand{\arraystretch}{1.2} 
    \resizebox{0.4\textwidth}{!}{ 
    \begin{ruledtabular}
        \begin{tabular}{lcc}  
            \textbf{Method} & \multicolumn{2}{c}{\textbf{Lattice parameters (Å)}} \\  
            & \textbf{$\alpha$-Ce} & \textbf{$\gamma$-Ce} \\  
            \hline
            \midrule
            LAK (this work) & 4.87 & 5.14 \\  
            OFR2 (this work) &4.62 & - \\
            $\mathrm{r^2SCAN}$ (this work) & 4.65 & -\\
            LDA \cite{casadei2016density} & 4.50 & - \\  
            PBE \cite{casadei2016density} & 4.68 & - \\  
            (EX+cRPA)@PBE0 \cite{casadei2012density} & 4.45 & 5.03 \\  
            Expt \cite{OLSEN1985129} & 4.83 & 5.16 \\  
        \end{tabular}
    \end{ruledtabular}
    }
    \caption{Comparison of lattice parameters of cerium from different functionals.}
    \label{tab:1}
\end{table}

Another interesting property of this phase transition is a change in magnetic behavior. The $\alpha$ phase is paramagnetic (i.e M=0) and  the $\gamma$ phase is ferromagnetic in nature. In Figure 2, we plot the magnetic moment (M) as a function of the lattice constant, using $\mathrm{r^2SCAN}$, OFR2 and LAK. No magnetic moment is shown in the $\alpha$ phase, but during the phase transition, there is a sudden jump in the magnetic moment. We obtained a magnetic moment of around 1.2 {$\mu_{B}$ per atom during the phase transition which is greater than results from other semi-local DFA's. \\
\begin{figure}[h]
    \centering
    \includegraphics[width=\linewidth]{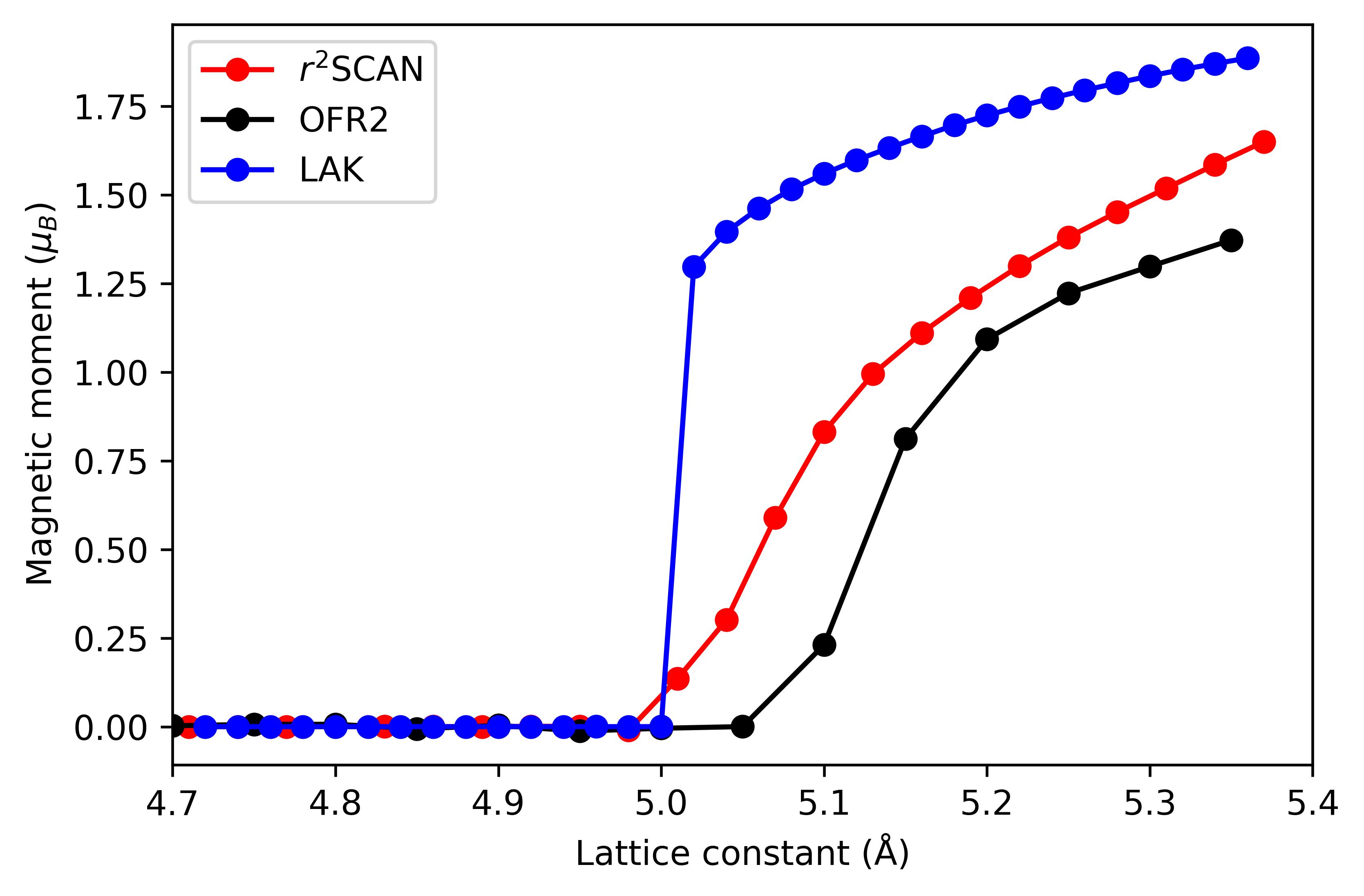}
    \caption{Variation of magnetic moment with lattice constant using different functionals} 
    \label{mag}
\end{figure}

We also calculated the critical pressure ($P_c$) for the same phase transition.
 The critical pressure is determined as the negative of the slope of the common tangent to the E-V curves of the $\alpha$- and $\gamma$-phases.
LAK predicts a critical pressure of -0.4 GPa, which is reasonably close to the extrapolated pressure (-0.8 GPa) to 0K \cite{casadei2016density}, and closer than the self-interaction-corrected spin polarized LDA calculations predicted in 1994 \cite{szotek1994self}. In contrast, (EX+cRPA)@PBE0 predicts -0.74 GPa, which is in closer agreement with the experiment. However, LAK predicts a volume collapse of 16\% at the transition, while (EX+cRPA)@PBE0 yields nearly 30\%.The experimental volume collapse at room temperature is approximately 15\%.}\\

\begin{table}[H]
  \centering
    \begin{tabular}{cccccc}
    \hline
    \hline
          &       & LAK   & Theoretical reference\cite{riseborough,casadei2016density}  &       & Expt.\cite{decremps2011diffusionless} \\
          \hline
     \textrm{$P_c$} (Gpa)     &       & -0.4  & -0.8  &       & 1.8, 1.44 \\
     \textrm{$\Delta V_{\gamma-\alpha}$} (\%)   &       & 15.8  & 28    &       & 14 \\
    \hline
    \hline
    \end{tabular}%
    \caption{Comparison of critical pressure ($P_c$) and volume collapse ($V_{\gamma-\alpha}$) during the phase transition.}
    \label{tab:lattice3}%
\end{table}%

\begin{figure}
    \centering
        \includegraphics[width=0.7\linewidth]{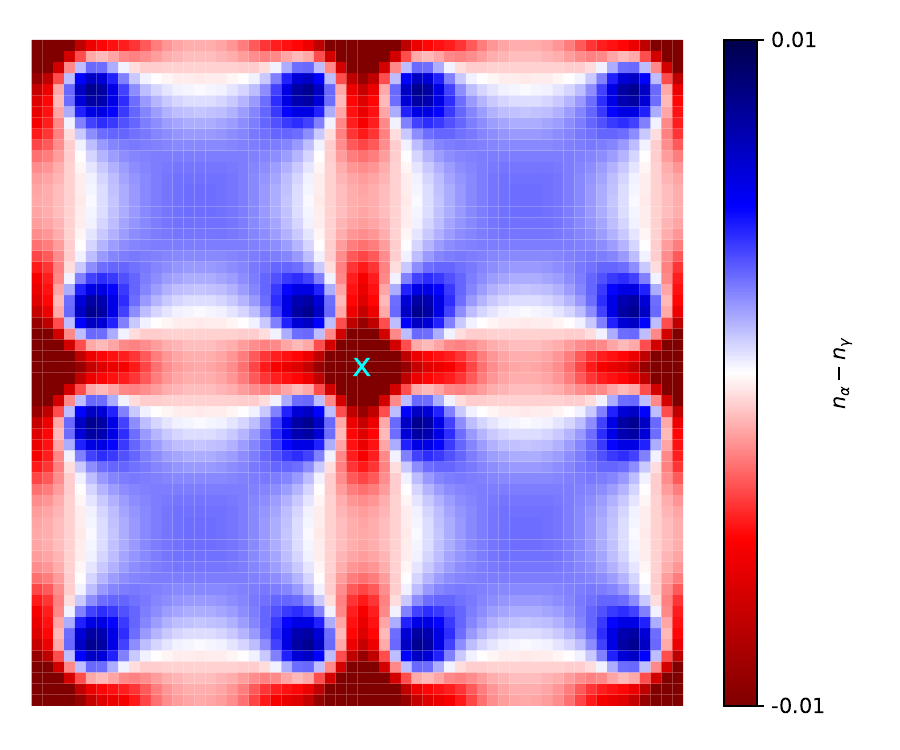}
        \caption{Projected density difference $n_\alpha - n_\gamma$ projected onto the $z=0$ plane. The density is averaged along $z-$axis at each $(x,y)$ grid point. The color scale spans from -0.01 to +0.01 $e/\mathrm{\AA}^3$. The atomic positions are in the deep red regions.}
        \label{fig4}
\end{figure}
    
Casadei et.al.\cite{casadei2012density} found that  $f$-electrons delocalize in the $\alpha$ phase and localize in the $\gamma$ phase. In order to investigate if semilocal LAK captures this localization/delocalization behavior across the phase transition, we plotted  the averaged the electron density difference, $n_\alpha-n_\gamma$, over the z-coordinate for a given x and y. Both phases were considered at the same volume of 125.51 $\AA^3$. The resulting plot, Fig. 3, shows more charge localization in the $\gamma$ phase at the atomic sites, with greater delocalization observed in the interstitial regions of the alpha phase. \\
\begin{figure}[t]
  \centering
  \begin{subfigure}[b]{0.48\columnwidth}
    \includegraphics[width=1.12\textwidth]{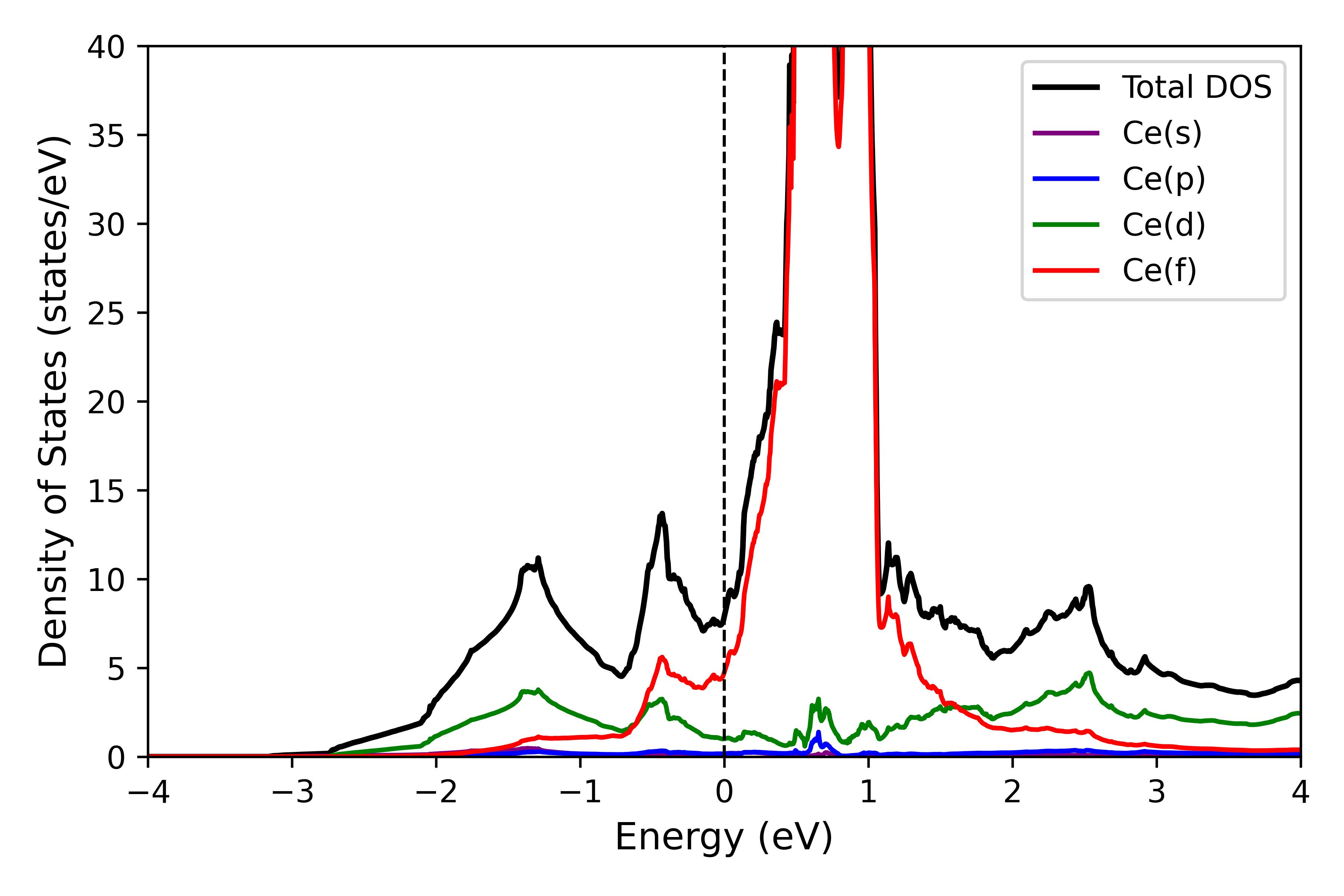}
    \caption{$\alpha$ phase}
    \label{fig:fig1}
  \end{subfigure}
  \hfill
  \begin{subfigure}[b]{0.48\columnwidth}
    \includegraphics[width=1.12\textwidth]{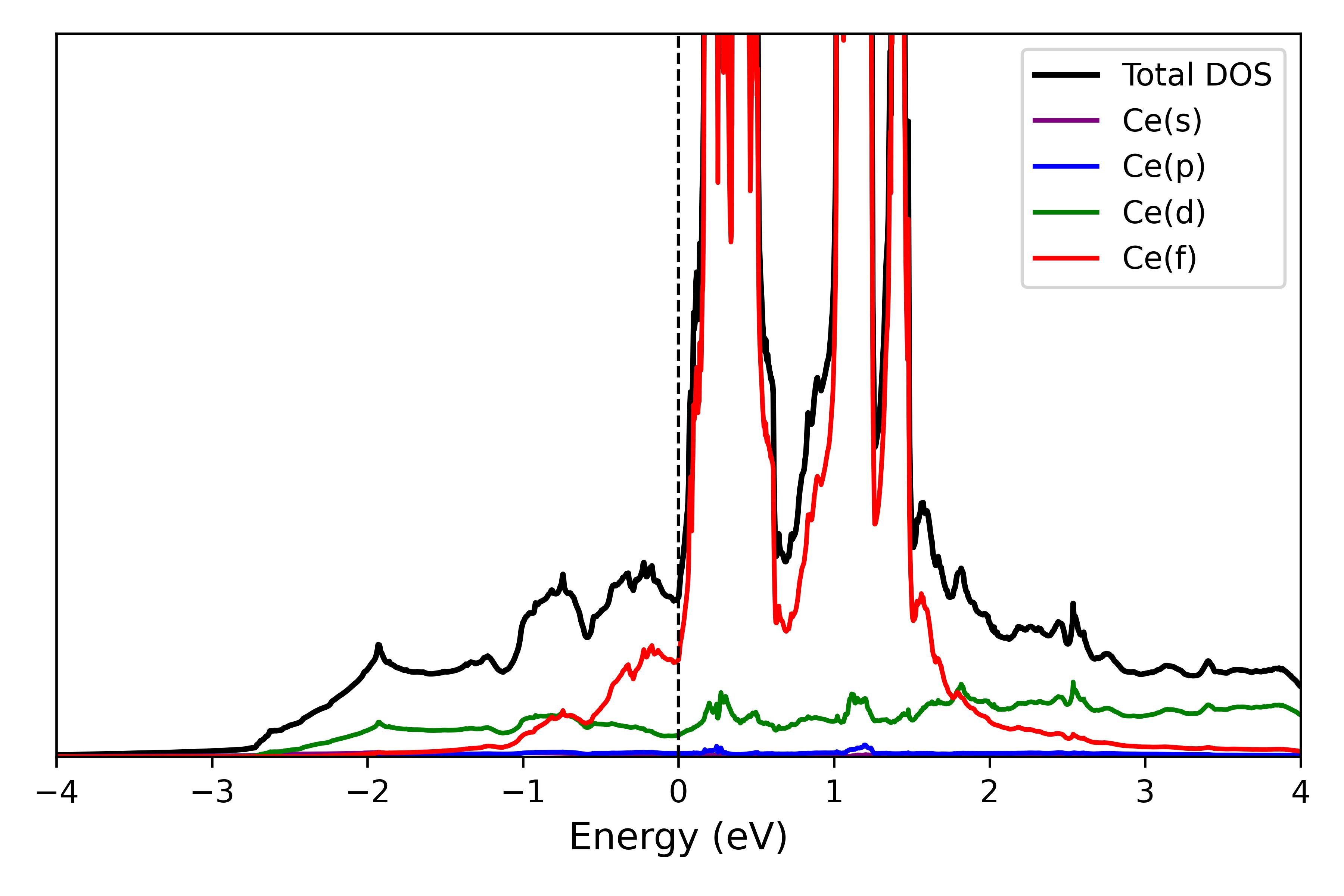}
    \caption{$\gamma$ phase}
    \label{fig:fig2}
  \end{subfigure}
  \caption{Density of States at equilibrium geometry of (a) 4.87 $\AA$ and (b) 5.14 $\AA$ respectively.}
  \label{fig4}
\end{figure}

We plotted the density of states (DOS) to see the contribution of $f$-electrons near the Fermi surface. The total DOS is the sum of spin-up and spin-down states.  Figure \ref{fig4} shows the total density of states with its orbital contribution. In both phases, we can see the major contribution, in total DOS, near Fermi surface is from $f$-electrons.It is noteworthy, that the DOS from LAK resembles the DOS from self-interaction-corrected LDA, producing a deep minimum of the unoccupied states above the Fermi level close to 1 eV \cite{szotek1994self}. In our work, the single f electron below the Fermi level seems to localize in the $\gamma$ phase without any dramatic changes in the orbital energy spectrum.

In summary, we have shown that the new meta-GGA LAK functional, with enhanced non-locality, can accurately predict both the $\alpha$- and $\gamma$-phases of cerium, as well as their correct energetic ordering at zero temperature. This is remarkable result from a single self-consistent calculation. The computed structural and magnetic properties agree well with the experimental results.The results are highly encouraging and demonstrate that meta-GGAs can become strong competitors of expensive fourth- and fifth-rung approximations at a fraction of their computational cost.\\
\textbf {Acknowledgments}:\\
The work of A.R was supported by Tulane University's startup funds.
A. G. and C. S. acknowledge support  by the Department of Energy Office of Basic Sciences under grant no. DE-SC0018331. We thank Prof. John P. Perdew for helpful discussions.

\bibliography{ref}
\end{document}